\documentstyle[aps,prd,preprint,tighten,amssymb,eqsecnum,newlfont]{revtex}
\newcommand{\beq}{\begin{equation}}
\newcommand{\eeq}{\end{equation}}
\newcommand{\bea}{\begin{eqnarray}}
\newcommand{\eea}{\end{eqnarray}}

\begin{document}
\preprint{
\begin{tabular}{r}
UWThPh-2000-07\\
January 2000
\end{tabular}
}
\draft
\title{On a possibility to determine the S--factor of the hep process
in experiments with thermal (cold) neutrons}
\author{
W.M. Alberico$^{\mathrm{a}}$,
S.M. Bilenky$^{\mathrm{b,a}}$ and
W. Grimus$^{\mathrm{c}}$
}
\address{$^{\mathrm{a}}$INFN, Sezione di Torino,\\
and Dipartimento di Fisica Teorica, Universit\`a di Torino,\\
Via P. Giuria 1, 10125 Torino, Italy}
\address{$^{\mathrm{b}}$Joint Institute for Nuclear Research, Dubna, Russia}
\address{$^{\mathrm{c}}$Institute for Theoretical Physics, University
of Vienna,\\
Boltzmanngasse 5, A--1090 Vienna, Austria}
\maketitle

\begin{abstract}
The problem of the magnitude of the S--factor of the hep process 
$p + {^3\mathrm{He}} \rightarrow {^4\mathrm{He}} + e^+ + \nu_e$
is one of the
most important unsolved problems in nuclear astrophysics. The magnitude of
$S(hep)$ has also an important impact on the interpretation of the
Super-Kamiokande and future SNO
solar neutrino results in terms of neutrino oscillations.
We point out the possiblity to determine the 
major hadronic contribution to $S(hep)$ from the
measurement of the total cross section of the process
$n + {^3\mathrm{H}} \rightarrow {^4\mathrm{He}} + e^- + \bar{\nu}_e$
in experiments with high intensity beams of thermal or cold neutrons.
\end{abstract}
\newpage
\renewcommand{\theequation}{\arabic{equation}}
The reaction
\beq
\label{main}
p + {^3\mathrm{He}} \to {^4\mathrm{He}} +e^+ + \nu_e 
\eeq
(hep reaction) is the source of solar neutrinos of the highest energies
(up to 18.8 MeV).
According to the Solar Standard Model (SSM) \cite{BP-95}, hep neutrinos
constitute only a negligible fraction of the solar neutrino flux:
the total flux of the hep neutrinos, predicted by the SSM, is equal to 
$2.1\times 10^3\, \nu$/cm$^2$\,s whereas the total SSM
flux of the main high energy $^{8}$B neutrinos with energy up to
15~MeV is equal to 
$5.15 \times 10^6\, \nu$/cm$^2$\,s \cite{BP-98}.

The recent interest in hep neutrinos 
(see Refs.\cite{BK-hep,fiorentini,escribano}) was triggered
by the results of the Super-Kamiokande experiment\cite{sol,nakahata,suzuki} 
in which the spectrum of the recoil electrons
in the solar neutrino-induced elastic neutrino -- electron scattering 
was measured. The spectrum measured by Super-Kamiokande collaboration
agrees with the predicted one in the whole energy region ($\geq 5.5$~MeV)
with the exception of the region of highest energies, 
in which a possible excess of events is observed.

The accuracy of the present data does not allow
to make definite conclusions on the distortion of the
recoil electron spectrum in the high energy region
(see Ref.\cite{suzuki}). However, the existing data can be considered 
as an indication in favour of such a distortion.
It was first shown by Bahcall and Krastev \cite{BK-hep} that, if the 
high-energy enhancement of the Super-Kamiokande spectrum is due to the
contribution of the hep neutrinos, then from the fit of the data it
follows that the value of the hep astrophysical S--factor $S(hep)$ is  
much larger than the SSM value 
\beq
S_{0}(hep)=2.3 \times 10^{-20}\: \mathrm{keV\, b} \,.
\label{Sfac}
\eeq

If there is no distortion of the $^8$B spectrum, then from the fit of 
the latest Super-Kamiokande data it was found \cite{suzuki} that 
$ S(hep)/S_0(hep)= 16.7$ ($\chi^2 = 19.5$ at 16 d.o.f.).
In the article of Bahcall \textit{et al.} \cite{BSK}, 
the  Super-Kamiokande data on the recoil electron spectrum were fitted
under the assumption of neutrino oscillations
with some representative values of neutrino oscillation parameters
$\sin^2 2\theta$ and $\Delta m^2$, taken in the corresponding 
allowed regions. The following ranges for $S(hep)$ were found: 
for the LMA MSW solution $27 < S(hep)/S_0(hep) < 47$,
for the SMA MSW solution $17 < S(hep)/S_0(hep) < 19$ and 
for the VO solution $0 < S(hep)/S_0(hep)< 40$.
A big progress in the investigation of the problem of the solar hep
neutrinos is expected in the near future: the accuracy of the measurement
of the electron spectrum in the Super-Kamiokande experiment 
will be increased \cite{suzuki} 
and the data of the new solar neutrino experiment SNO \cite{SNO}
will appear. In the SNO experiment the spectrum of
solar $\nu_{e}$'s will be determined from the measurement of the electron
spectrum in the process $\nu_e + d \to e^{-} + p + p$.

One expects that the cross section of the hep process 
at the solar energies ($\lesssim 30$~keV)
is very small and cannot be directly measured in an experiment. 
The most detailed calculations of the cross section of the
hep process have been done by Carlson {\em et al.}\cite{Carl} and by
Schiavilla {\em et al.}\cite{Schiav}. According to 
these calculations, the value of the hep cross section is
$\sigma(hep) \simeq 6\times 10^{-51}$ cm$^2$ at $E = 20$~keV,
where $E$ is the kinetic energy of the initial particles in the C.M. system.
The value (\ref{Sfac}) of the astrophysical S--factor  
obtained in Ref.\cite{Schiav} is adopted in the SSM \cite{BP-95}.

The cross section of the hep process calculated in \cite{Schiav}
is strongly suppressed by two reasons. The first reason is connected
with SU(4) symmetry: in the
allowed approximation the matrix element of the process vanishes if
only the dominant $s$--states
of the $^4$He and $^3$He wave functions are taken into account\cite{werntz}.
The second reason lies in the cancellation 
between the matrix element of the usual one--body hadronic current and the
matrix elements of the (two--body) $\pi$ and $\rho$ exchange currents, which
also account for the transition of nucleons to intermediate $\Delta$ isobar
states.
 
The problem of the cross section of the hep process is considered as one
of the most important problems of nuclear astrophysics \cite{Bahcall}.
There are many uncertainties in the existing (model-dependent and complicated)
 calculations of this cross section.
It is very desirable to find a way to obtain an {\it experimental}
information on the astrophysical S--factor of the hep process.
The aim of this letter is to propose such an experiment by exploiting the fact
that the hadronic matrix element of the process 
$n + {^3\mathrm{H}} \to {^4\mathrm{He}} + e^- + \bar\nu_e$ is equal to that of
the hep process due to isotopic invariance of strong interactions.

The total cross section of a process with charged initial particles 
at small energies has the form \cite{BahcAdel}
\begin{equation}
\sigma (E)=\frac{1}{E}\, S(E)\, e^{-2\pi \eta} \,.
\label{sighep}
\end{equation}
Here $E $ is the kinetic energy of the initial particles in the C.M. system
and
\begin{equation}
\eta =\frac{Z_1 Z_2 e^2}{v} \,,
\end{equation}
where $Z_1e$ and $ Z_2e$ are the charges of the initial particles and
$ v=\sqrt{2E/\mu}$
is the relative velocity with the reduced mass $\mu$.
The  quantity $ P = e^{-2\pi \eta}$ in the relation (\ref{sighep}) 
is the Coloumb penetration
factor and the S--factor $S(E)$ is determined mostly by the strong
interactions. If there are no resonances at small energies the S--factor
depends very weakly on $E$ \cite{Fowler,BahcAdel}. 

One of the reasons of the smallness of the nuclear cross section at small
solar energies is the suppression due to the penetration factor.
For example, for the hep process at $E = 15 $~keV  one obtains 
$P \simeq 8 \times 10^{-7}$ and at 
$E = 20$~keV, $P \simeq 5 \times 10^{-6}$.

Let us now compare the process
\begin{equation}
n + {^3{\mathrm{H}}} \rightarrow {^4{\mathrm{He}}} + e^- + \bar{\nu_e} \,,
\label{ntritium}
\end{equation}
for which obviously there is no Gamov penetration factor, with the hep process
(\ref{main}). From isotopic SU(2) invariance of the strong interactions it
follows that the hadronic parts of the matrix elements of the hep process
and of the process (\ref{ntritium}) are the same. 
In fact, 
with 
$J^{(\pm)}_{\alpha} \equiv J_{\alpha}^{1 \pm i2} \equiv 
J^1_\alpha \pm iJ^2_\alpha$ being 
the ``minus'' and ``plus'' components, respectively, 
of the isovector hadronic V--A current in
the \emph{Heisenberg representation},
we obtain
\beq
\langle {^4\mathrm{He}}|J^{(-)}_\alpha|p\, {^3\mathrm{He}} \rangle =
\langle {^4\mathrm{He}}|{\cal U}^{-1}{\cal U}J^{(-)}_\alpha
{\cal U}^{-1}{\cal U}|p\, {^3\mathrm{He}} \rangle
= - \langle {^4\mathrm{He}}|J^{(+)}_\alpha|n\, {^3\mathrm{H}} \rangle \,,
\label{charsym}
\end{equation}
where $ {\cal U}=\exp(i\pi T_2 )$ is the operator of rotation by an
angle  $\pi$ around the second axis in the isotopic space.
In $J^{(\pm)}_{\alpha}$ all strong interactions are taken into account. 
With the help of the relation (\ref{charsym})
we can connect the cross section of the process (\ref{ntritium})
with the astrophysical S--factor of the hep process.

In the region of very small energies we are interested in, 
only the $s$--wave of the initial particles is relevant and the
hadronic matrix element
$\langle {^4{\mathrm{He}}}|J^{(-)}_{\alpha}(0)|p\,{^3{\mathrm{He}}}\rangle$
is constant. For the hep cross section one has to take into account the Coulomb
interaction of the initial $p$ and $^{3}$He by the usual Coulomb factor
\begin{equation}
\frac{|\psi^{(+)}_{\vec{p}}(0)|^2}
{|\psi_{\vec{p}}|^2}=\frac{2\pi \eta}{e^{2\pi \eta}-1}
\simeq 2\pi \eta\, e^{-2\pi \eta} \,,
\label{Coul}
\end{equation}
where the approximate expression holds for small C.M. kinetic energies $E$. 
With Eq.(\ref{Coul}) we obtain the cross section 
\begin{equation}
\sigma(hep)=\frac{(2\pi)^7}{E}\, G^2_F\, e^2\, m_e^5\, \mu \sum_\mathrm{spins}
\left|\langle{^4{\mathrm{He}}}|\vec{J}\,^{(-)}|p\,{^3{\mathrm{He}}}\rangle
\right|^2 f(-2,\varepsilon_{0})\, e^{{-4\pi e^2}/{v}} \,.
\label{sighep2}
\end{equation}
Here $m_e$ is the electron mass and 
\begin{equation} 
f(Z,\varepsilon_{0})=\int^{\varepsilon_{0}}_1
F(Z,\varepsilon)(\varepsilon_{0} - \varepsilon)^{2}
\sqrt{\varepsilon^{2}-1}\,\varepsilon d\varepsilon
\label{fsmall}
\end{equation}
is the phase space factor,  $\varepsilon ={k_0}/{m_e} $ and 
$\varepsilon_{0}={\Delta}/{m_e}$ with \cite{firestone}
\begin{equation}\label{deltahep}
\Delta = m_p + m_{^3{\mathrm{He}}} - m_{^4{\mathrm{He}}} 
\simeq 19.284~{\mathrm{MeV}}.
\end{equation}
The Fermi function $F(Z,\varepsilon)$ with $Z=-2$ takes into account 
the Coulomb interaction of the final $e^{+}$ and $^4$He.
In our case, this function is practically equal to one due to the large
positron energies up to 19 MeV (\ref{deltahep}).
Notice that in the process (\ref{main}) the isotopic spin is
changed by one and only the Gamov--Teller transition must be taken
into account (for the details of the derivation of (\ref{sighep2})
see Ref.\cite{ABBG}).

In essence, taking into account the relation (\ref{charsym}),
the difference between the hep process (\ref{main}) and
the process (\ref{ntritium}) is given by the Coulomb correction factor
(\ref{Coul}). Therefore, in the region of small energy we 
have the following expression for the ratio of the cross sections of the
process (\ref{ntritium}) and the hep process:
\beq
\frac{\sigma(n\, {^3\mathrm{H}}\to {^4{\mathrm{He}}}\: e^- {\bar\nu}_e)}%
{\sigma(hep)} = 
\frac{1}{2\pi \eta\, e^{-2\pi \eta}} \, 
\frac{f(2, \varepsilon'_0)}{f(-2, \varepsilon_0)} \,.
\label{sigtri}
\eeq
In this relation we have also taken into account 
the difference in the phase space
factors. As in the hep process, the Fermi function can be well approximated
by one, however, the isospin-breaking effect due to the difference between
$\Delta$ (\ref{deltahep}) and (see Ref.\cite{firestone})
\beq\label{delta'}
\Delta' = m_n + m_{^3H} - m_{^4\mathrm{He}} \simeq 21.107~\mathrm{MeV}
\eeq
cannot be neglected. The reason is that 
$f(-2, \varepsilon_0) \simeq \varepsilon_0^5/30$ and analogously for
$f(2, \varepsilon')$ with $\varepsilon' = \Delta'/m_e$ and, therefore, 
the energy releases (\ref{deltahep}) and (\ref{delta'}) 
enter with the 5th power.
Now using the  relation (\ref{sighep}) which defines the S--factor,
we obtain the following
relation between the cross section of the process (\ref{ntritium})
and the astrophysical S--factor of the hep process:
\begin{equation}
\sigma(n\, {^3\mathrm{H}}\to {^4{\mathrm{He}}}\: e^- {\bar\nu}_e) \simeq
\frac{S(hep)}{2\pi e^2\sqrt{2\mu E}} 
\left( \frac{\Delta'}{\Delta} \right)^5 \,.
\label{sigtri2}
\end{equation}
From the above relation we can determine  $S(hep)$
if the cross section of the process
$n\, {^3\mathrm{H}}\to {^4{\mathrm{He}}}\: e^-\, {\bar\nu}_e$ 
will be measured. From Eq.(\ref{sigtri2}), using 
$(\Delta'/\Delta)^5 \simeq 1.57$, we arrive at the numerical estimate
\beq
\sigma(n\, {^3\mathrm{H}}\to {^4{\mathrm{He}}}\: e^-\, {\bar\nu}_e) \simeq
1.3\times 10^{-47}\, \frac{1}{\sqrt{E[\mathrm{MeV}]}}
\frac{S(hep)}{S_0(hep)}\;\mathrm{cm}^2 \,,
\label{numeric}
\eeq
where $S_0(hep)=2.3\times 10^{-20}$~keV~b is the value of $S(hep)$
adopted in the SSM.

If we put $ S(hep)=S_0(hep)$, for the cross section of the process
(\ref{ntritium}) we obtain the following estimates 
for thermal ($E\simeq 2.5\times 10^{-2}$~eV), cold
($E\simeq 2\times 10^{-3}$~eV) and ultracold ($E\simeq 2\times 10^{-4}$~eV)
neutrons \cite{West}:
\beq
\sigma(n\, {^3\mathrm{H}}\to {^4{\mathrm{He}}}\: e^-\, {\bar\nu}_e)
\simeq 10^{-43} \: \mathrm{cm}^2 \times
\left\{ \begin{array}{c} 1.3 \\ 4.7 \\ 15 \end{array} \right\}
\; \mbox{for} \:
\left\{ \begin{array}{c} \mathrm{thermal} \\ \mathrm{cold} \\
\mathrm{ultracold} \end{array} \right\} \: \mathrm{neutrons} \,.
\eeq

In conclusion, from isotopic SU(2) invariance of strong interactions
we have obtained a relation that connects the main hadronic part of
the astrophysical S--factor of the hep process (\ref{main})
with the total cross section of the process 
$n + {^3\mathrm{H}} \rightarrow {^4\mathrm{He}} + e^- + \bar{\nu}_e$
at low energies. To work with a $^{3}$H target is a difficult 
problem.\footnote{For a liquid tritium target see the paper of 
H.H. Grafov \textit{et al.} \cite{Grafov}.}
But there are several advantages in the investigation of the 
$n\,{^3\mathrm{H}}$ process (\ref{ntritium}):
\begin{enumerate}
\item
There is no Coloumb penetration factor which strongly suppresses the
cross sections of processes with equally charged particles at very small
energies.
\item
The matrix element of this process determines the main hadronic part of
the S--factor of the hep process. This follows from simple isotopic
symmetry of strong interactions. Note that this is not the case for the
reaction $n + {^3\mathrm{He}} \to {^4\mathrm{He}} + \gamma$ (see, for 
example, Ref.\cite{Carl}).
\item
The possibility of using high intensity very low energy
thermal, cold and  ultracold (?) neutron beams is a unique possibility
to increase the cross section of the weak process due to the famous
$1/v$ effect.
\item
The signature of the process (\ref{ntritium}) are electrons with
energies up to 20~MeV, which is a much higher energy than the one of
the background electrons from the decay of $^{3}$H.
\end{enumerate}

\section*{Note added}

After this paper was finished and sent to the editor, new calculations
of the hep cross section appeared. In 
L.E. Marcucci \textit{et al.}, nucl-th/0003065 and nucl-th/0006005, 
it is claimed that, inspite
of the centrifugal suppression, the $p$--state of the initial $p$ --
$^3$He system gives a sizable contribution to the cross section of the
hep process (see also C.J. Horowitz, Phys. Rev. C \textbf{60}, 022801
(1999)). The value of $S(hep)$ that was obtained by 
L.E. Marcucci \textit{et al.} is approximately 5 times larger than the
value found in Ref.\cite{Schiav} and ascribes to the $p$--wave 
contribution about $40\%$ of the evaluated cross section. 
On the other hand, in T.-S. Park \textit{et al.}, nucl-th/0005069, 
based on an effective field theory of QCD, no explicit role is
attributed to the $p$--wave contribution.

Our paper is based on the classical assumption that in the region of
small energies (keV) of the proton only the $s$--wave is
important (see, e.g., D.D. Clayton, \textit{Principles of stellar
evolution and nucleosynthesis} (McGraw-Hill, New York, 1968)). 
Should the importance of the $p$--wave be confirmed by
further calculations, the method we proposed here allows to determine
a lower bound on $S(hep)$ which is given by the major $s$--wave
contribution. Finally we want to stress that a measurement of the
process 
$n + {^3\mathrm{H}} \rightarrow {^4\mathrm{He}} + e^- + \bar{\nu}_e$
at thermal neutron energies would allow to test theories of nuclear
reactions, which are relevant for astrophysics.

In  conclusion, we would like to stress that, in spite of the fact that
the expected cross section of the above $n + {^3}$H process with
thermal neutrons is relatively large, the proposed experiment is a very
difficult one. A basic
requirement for such an experiment is the availability of a large tritium 
target. We would like to notice that 
a 4 kg $^3$H target exists and was discussed in the poster session P25
(L.N. Bogdanova \textit{et al.})
at the recent conference Neutrino 2000 in Sudbury, June 2000.
Of course, the problem of background
to the process we propose is very severe. Let us notice that in order to
reduce the background from cosmic rays, one could conceive to perform the
experiment at a powerful underground reactor, like the one in
Krasnoyarsk.

\end{document}